\documentstyle[12pt]{article}
\begin{document}
\begin{center}
\Large \bf Heavy Ion Collisions and the Density Dependence
of the Local Mean Field\footnote{Supported
by BMFT(06T\"U746(2)) and by GSI(T\"U F\"as T).} \\ \vspace*{5mm}
\normalsize
 \vspace*{9mm}
\large T. Kubo, E. Lehmann, Amand Faessler,
Rajeev K. Puri\footnote{present address:
SUBATECH, Laboratoire de Physique Nucleaire,
Universite de Nantes, 2, rue de la Houssiniere, F-44072 Nantes Cedex 03,
France},\\
N. Ohtsuka\footnote{present address: Management Information Department,
Onomichi Colledge, Onomichi, 722 Hiroshima, Japan},
K. Tsushima, J. Zipprich and S.W. Huang
\\ \normalsize
\vspace*{6mm}
Institut f\"ur Theoretische Physik, Universit\"at T\"ubingen,\\
Auf der Morgenstelle 14, 72706 T\"ubingen, Germany
\\ \vspace*{5mm}
\today \\ \vspace*{5mm}
\end{center}
\vspace*{12mm}
\it Abstract: \\ \\
\normalsize
\baselineskip 24pt
We study the effect of the density dependence of the scalar and the
vector part of the nucleonic self-energy in
Relativistic Quantum Molecular Dynamics (RQMD) on observables like
the transversal flow and the rapidity distribution.
The stability of nuclei in RQMD is greatly improved if the density
dependence is included
in the self-energies compared to a calculation assuming always saturation
density of nuclear matter.
Different approaches are studied: The main results are calculated with
self-energies extracted from a Dirac-Br\"uckner-Hartree-Fock G-matrix of
a one boson exchange model, i.e. the Bonn potential.
These results are compared with those obtained by a generalization of
static Skyrme force, with calculations in the simple linear Walecka model
and results of the Br\"uckner-Hartree-Fock G-matrix of the Reid soft core
potential. The transversal flow
is very sensitive to these different approaches. A comparison with
the data is given.
\\
\begin{center}
\large \bf 1. Introduction
\end{center} \normalsize
\baselineskip 24pt
The dynamics of heavy-ion reactions depends on the mean field and
the description of binary collisions.
Hence, the mean field and the cross sections
for binary collisions are the main ingredients of transport models like
Quantum Molecular Dynamics (QMD) \cite{Aichelin91,Aichelin86}
and the BUU-type models \cite{Kruse85,Bonasera89,Gregoire87,Bauer86,Bertsch88,
Bertsch84} which
are proven to describe successfully the dynamics of heavy-ion collisions.
Finally, both ingredients should be determined from the same parameter free
microscopic force. The strong repulsive short range correlations of these
forces
necessitate a non-pertubative self-consistent treatment as done in Br\"uckner
theory.

In the non-relativistic regime such a self-consistent description of heavy-ion
reactions was done in the framework of QMD\cite{Jaenicke92,Khoa92}. In this
approach self-consistent
potentials as well as in medium cross sections are extracted from
non-relativistic
G-matrix elements which are given by Br\"uckner-Hartree-Fock (BHF)
calculations.
In these calculations the Bethe-Goldstone equation including an angle
averaged Pauli operator was solved for the situation of two pieces of nuclear
matter penetrating each other. As a bare nucleon-nucleon interaction the
Reid-soft-core potential is used in this approach.

In the covariant extension of QMD, i.e. Relativistic QMD (RQMD)
\cite{Sorge89,Maruyama91a,Lehmann95}, up to now only
static Skyrme forces combined with the so-called Cugnon parametrization of the
nucleon-nucleon cross section were used. These generalized Skyrme forces
are treated
as scalar potentials in the framework of Constrained Hamilton Dynamics.
But the full
Lorentz-structure of the nucleon-nucleon interaction contains large scalar and
vector components. Effects of these relativistic forces were already studied in
the
framework of covariant generalizations of models of the BUU-type
\cite{Ko87,Cassing90b,Fuchs92}, mostly in the framework of the
$\sigma -\omega$ model, but not yet in RQMD.

In order to determine the scalar and vector components of the self-energy of
the
nucleon from realistic forces in a covariant fashion the solution of the Bethe-
Salpeter equation with Pauli-operator is needed, which can be achieved in a
three
dimensional reduction of the Bethe-Salpeter equation as given by
the Thompson equation. Nuclear matter calculations
\cite{Brockmann84,terHaar87} in this framework are done by
using one boson exchange potentials (OBEP), e.g. the Bonn
potential, as the bare nucleon-nucleon interaction.
The self-energy components extracted from these Dirac-Br\"uckner-Hartree-Fock
(DBHF) calculations show a strong density dependence while the momentum
dependence
is moderate.
Therefore, in the present approach we like to study as a first step the density
dependence of the self-energy components in a local density approximation
(LDA).
For this purpose we use in RQMD simulations self-energy components as extracted
from G-matrix elements determined by DBHF calculations using the Bonn potential
as well as
a parametrization \cite{Elsenhans92} footing on BHF calculations using the
Reid-soft-core potential.
In both approaches we determine observables which reflect on the time evolution
of the
phase space, namely, the rapidity distribution and the transversal flow inside
the
reaction plane.

In the second chapter we sketch our RQMD approach. In addition, in this chapter
we give shortly an outline how the results extracted from
the Br\"uckner approach
for the mean field using the solutions of the non-relativistic Bethe-Goldstone
equation
and the relativistic Bethe-Salpeter equation as well are used within RQMD in
a local density approximation.
In the third chapter we discuss the results and compare them with different
approaches and data.
In the last chapter we finally summarize our main conclusion.

\newpage
\begin{center}
\large \bf 2. Density Dependent Interactions within RQMD
\end{center} \normalsize
\baselineskip 24pt

In this section we give a brief description of the formalism of
relativistic Quantum Molecular Dynamics (RQMD)
with relativistic self-energies.

In the formalism of Constrained Hamiltonian Dynamics \cite{Dirac49,Samuel82},
the original
8N dimensional phase space of
N interacting relativistic particles is reduced to usual 6N dimensions by 2N
constraints.  These constraints fix the individual energies and
parametrize the world lines by fixing the relative time
coordinates.

The energies are fixed by N on-shell conditions
\begin{equation}
K_{i}=(p^{\mu}_{i}-\Sigma^{\mu}_{i})(p_{i\mu}-\Sigma_{i\mu})
-(m_{i}-\Sigma^{(s)}_{i})^{2}=0,
\end{equation}
which contain scalar and vector potentials.  Hence, the constraints (1) take
the full
Lorentz structure of the $NN$ interaction into account explicitly.
We regard the self-energy components $\Sigma^{(s)}_{i}$ and
$\Sigma^{\mu}_{i}$
as sums of two-body mutual interactions-at-a-distance, i.e.
$\Sigma^{(s)}_{i}=\sum_{j\neq i}\Sigma^{(s)}_{ij}$ and
$\Sigma^{\mu}_{i}=\sum_{j\neq i}\Sigma^{\mu}_{ij}$.
In order to use the self-energies provided by relativistic $NN$ interactions,
we adopt a kind of local density approximation (LDA).
In the relativistic Hartree
approximation the self-energies are proportional to the scalar density
$\rho_{s}$ and the baryon current $B^{\mu}$, i.e.
\begin{equation}
\Sigma^{(s)}=\Gamma_{s}\rho^{s},\hbox{   }
\Sigma^{\mu}=\Gamma_{v}B^{\mu}.
\end{equation}
Interpreting the sum of the Lorentz scalar two-body interaction densities
\begin{equation}
\rho_{ij}=\frac{1}{(4\pi L)^{3/2}}exp(q^{2}_{Tij}/4L),
\end{equation}
as a scalar density $\rho^{s}_{i}=\sum_{j \neq i}\rho_{ij}$
one can define in addition the baryon current
in a consistent way and gets the self-energies of a particle $i$ as
\begin{equation}
\Sigma^{(s)}_{i}=\Gamma_{s}\sum_{j \neq i}\rho_{ij}, \hbox{     }
\Sigma^{\mu}_{i}=\Gamma_{v}\sum_{j \neq i}\rho_{ij}u^{\mu}_{j}.
\end{equation}
Here
$q^{\mu}_{Tij}=(g^{\mu\nu}-\hat{p}^{\mu}_{ij}\hat{p}^{\nu}_{ij})q_{ij\nu}$,
$q^{\mu}_{ij}=q^{\mu}_{i}-q^{\mu}_{j}$, $p^{\mu}_{ij}=p^{\mu}_{i}+p^{\mu}_{j}$
, $\hat{p}_{ij}=p_{ij}/(p^{2}_{ij})^{1/2}$ and $u^{\mu}_{j}$ being the
4-velocity
of particle $j$.
In this sense the self-energies are given by sums of two-body mutual
interactions-at-a-distance.  The dependence of these self-energies on
$q^{2}_{Tij}$ ensures the multi-time description used in RQMD without
further additional approximations. A detailed discussion proving the
reliability of these
definitions is given in the appendix.

In the case of realistic $NN$ interactions the strong short-range
components require to take account for the effects of two-body correlations.
This can be attained in the framework of Br\"uckner theory.
Results of self-consistent Dirac-Br\"uckner calculations can be used
in the approach discussed above with help of effective coupling functions
$\Gamma_{s}$ and $\Gamma_{v}$, which are density dependent.
This density dependence is usually not included.
We shall show that the inclusion of this density dependence improves the
description greatly.
In this way, realistic forces extracted from Dirac-Br\"uckner
calculations can be implemented in RQMD.

For the actual simulations we have used 4 kinds of self-energies extracted from
the following
models and theories:

(1) Skyrme model for the soft equation of state \cite{Maruyama91a, Lehmann95}
(Skyrme),

(2) $\sigma$-$\omega$ model \cite{Walecka74,Serot88} (QHDI),

(3) Br\"uckner Hartree Fock theory with Reid soft core potential
\cite{Elsenhans92} (BHF),

(4) Dirac-Br\"uckner Hartree Fock theory with Bonn potential \cite{Malfliet94}
(DBHF).

In order to obtain fully covariant constraints, we have fitted the density
dependent coupling
constants as functions of the local scalar density to reproduce the original
calculations of scalar
and vector self-energies of models (3) and (4) by means of the Lagrange
interpolation
method.  The quality of the fit is shown in Fig.1.

We have followed the ref.\cite{Lehmann95} for the choice of the other
additional $N$ constraints
to specify the world lines of the particles:
\begin{equation}
\chi_{i}=\sum_{j \neq i}w_{ij}p^{\mu}_{ij}q_{\mu ij}=0,\hbox{   }i=1,...,N-1
\end{equation}
\begin{equation}
\chi_{N}=\hat{P}^{\mu}Q_{\mu}-\tau=0
\end{equation}
with $\hat{P}^{\mu}=P^{\mu}/(P^{2})^{1/2}$, $P^{\mu}=\sum_{i}p^{\mu}_{i}$,
$Q^{\mu}=\sum_{i}q^{\mu}_{i}/N$ and the dimensionless scalar weight function
\begin{equation}
w_{ij}=\frac{1}{q^{2}_{ij}/L_{c}}exp(q^{2}_{ij}/L_{c})
\end{equation}
with $L_{c}=8.66$ fm$^{2}$ and $\tau$ denotes a global time evolution
parameter.

The total Hamiltonian is now expressed as
\begin{equation}
H=\sum_{i=1}^{2N-1}\lambda_{i}\phi_{i}
\end{equation}
where $\phi_{i}=K_{i}$ for $i \leq N$ and $\phi_{i}=\chi_{i-N}$ for $N+1 \leq i
\leq 2N-1$,
then the equations of motions are given by
\begin{equation}
dq^{\mu}_{i}/d\tau=[H,q^{\mu}_{i}],\hbox{   }dp^{\mu}_{i}/d\tau=[H,p^{\mu}_{i}]
\end{equation}
where $[A,B]$ means the Poisson bracket of phase space functions $A$ and $B$.
The evolution of the system can be computed by integrating the set of above
equations (9),
but we have to determine the unknown Lagrange multipliers $\lambda_{i}(\tau)$
to calculate
the dynamics.
This can be done using the fact that the complete set of $2N$ constraints
$\phi_{i}$
 $(\phi_{2N}=\chi_{N})$ must be fulfilled during the whole time evolution, i.e.
\begin{equation}
d\phi_{i}/d\tau=\partial \phi_{i}/\partial\tau + [H,\phi_{i}]=0.
\end{equation}
Using the Hamiltonian (8), one gets for the Poisson bracket
\begin{equation}
[H,\phi_{i}]=\sum_{j=1}^{2N-1}\lambda_{j}[\phi_{j},\phi_{i}]
\end{equation}
thus inverting the matrix $C_{ji}=[\phi_{j},\phi_{i}]$, we obtain the
multipliers as
\begin{equation}
\lambda_{i}=C^{-1}_{iN}.
\end{equation}

If the Dirac's first class condition is fulfilled, i.e. $[K_{i},K_{j}]=0$,
(in our case the condition has been confirmed numerically during the whole time
evolution.)
the Hamiltonian is reduced into a simpler form:
\begin{equation}
H=\sum_{i=1}^{N}\lambda_{i}K_{i}
\end{equation}
\begin{equation}
\lambda_{i}=\Delta^{-1}_{iN}
\end{equation}
where $\Delta_{ji}=[K_{j},\chi_{i}]$ is a submatrix of the constraint matrix
$C_{ji}$.

For more details, especially for the foundations of these
formalisms, we refer to \cite{Aichelin91} in case of QMD and to
\cite{Sorge89} and \cite{Lehmann95} in case of RQMD.
We note that the binary collisions are dealt in RQMD in the same way as in
the non-covariant QMD by using Monte Carlo methods.

\begin{center}
\large \bf 3. Results
\end{center} \baselineskip 24pt

First we compare the stability of RQMD nuclei described with different
interactions
obtained from the above mentioned 4 models and theories.  It is important to
check the stability
at actual boosting energies, since the momentum dependence of the self-enegies
hides the
trivial analytic expressions of covariance.  We adopt the energy at $1.8$ GeV/A
for later simulations.
Figure 2 shows the root mean square radii of Ca versus time, with 6 different
approaches:
Figure 2 contains the time dependence of phenomenological models, i.e. static
Skyrme and linear Walecka
models.  As is easily seen, although the Skyrme model works well, but in the
simple linear
Walecka model the nuclei dissassemble
after several time steps.  This is understandable: the strong constant
couplings
directly reflect on strong repulsive components of the force when the local
density reaches
high values
because of the Fermi motion of the nucleons, however it might be expected that
the density
dependence of the self-energies recovers the discrepancy, since the
self-energies are
softened in high density domain if the two body correlations are included in
Br\"uckner theory
(see Fig.1), thus the correlation, i.e. density dependence of the couplings
would give a counter balance automatically inside of nuclei.
In order to visualize this mechanism clearly, thus we plot the time dependence
with microscopic
density dependent self-energies and the self-energies at the saturation point
from BHF and DBHF
calculations in Figs.2.
One can see that the density dependence of the self-energies strongly improved
the stability of
the nuclei.

One of the most interesting physical observable in heavy ion collision at
intermediate energies are
rapidity distributions and the collective transverse flow.  These quantities
are very sensitive
to the interactions used in the models,
thus frequently used to investigate the nuclear equation of state far from
the saturation point.
In order to be able to compare with the experimental data,
we have performed simulations of the nearly symmetric collision Ar+KCl at $1.8$
GeV/A
which were measured with the streamer chamber at the BEVALAC at LBL
\cite{Stroebele83}.
In the calculation we are using the above 3 kinds of stable nuclei, i.e.
simulated by
(1) Skyrme (2) BHF (3) DBHF self-energies
and compared with the experimental data \cite{Danielewicz85}.
The initialization of the target and the projectile has been carried out by
standard Monte Carlo methods as described in ref.\cite{Aichelin91}.
According to the ref.\cite{Danielewicz85}, we have chosen the impact parameter
range to be
$0 \leq b \leq 2.3$ fm.  The standard Cugnon parametrization is used for the
binary
nucleon-nucleon cross sections \cite{Cugnon88}.

The collective transverse flow obtained from our simulation is plotted in
Fig.3.
It is quite remarkable that the conventional Skyrme force (soft EOS) is not
able to reach the experimentally measured magnitude of the flow distribution at
this energy. The reason is clear: The force contains only static interactions
between
constituent particles and no explicit energy dependence.
However, on the contrary, the experimental data lie in between DBHF and BHF
results.  The
Schr\"odinger equivalent potential extracted from the DBHF self-energies shows
a linear
energy dependence in our approach.
Hence one gets a repulsive interaction at high relative momenta which produce a
strong sidewards flow, which is reflected on the $P_{x}/A$.
There is still deviation between the theory and the experiment, it might be
overcome
if one takes the configuration dependence (deviation of the momentum
distribution from a Fermi sphere) in the momentum space of the selfenergies
into account,
means by using local configurations which can be parametrized by
two penetrating fermi spheres \cite{Elsenhans92}.

The rapidity distributions is also shown in Fig.4.
The asymmetric property of the distribution is due to the system we study.
One can notice that the Skyrme result exhibits less and BHF gives the highest
stopping,
but the differences between the various approaches are rather small.
We found that the Skyrme provides a little bit less Pauli blocking (around 10
percent
compared with the realistic forces.).  And also, the stability of single
nuclei might affect on this quantity, since it is not quite stable using
BHF in comparison when using static Skyrme interactions or interactions
extracted from DBHF (see Fig.2).


\begin{center}
\large \bf 4. Conclusions
\end{center}\baselineskip 24pt
We investigated the effect of realistic forces in comparison to
phenomenological forces on the sidewards flow and the rapidity
distribution created during heavy-ion reactions by using the
covariant RQMD approach. The realistic forces are extracted from
Dirac-Br\"uckner calculations and non-relativistic Br\"uckner calculations
as well while the phenomenological forces are determined by
local static Skyrme interactions or in the framework of a simple
linear Walecka model.

It was proven that single nuclei are stable during the time span
needed for a simulation of heavy-ion collisions if the full density
dependence (beyond a linear one) of the interaction is included when using
realistic forces.

The rapidity distribution at the final state shows that
realistic interactions create slightly more stopping than the
static Skyrme interaction.

Due to the energy dependence and the correct density dependence
of these realistic forces the flow is enhanced in comparison to
calculations using static Skyrme interactions. Furthermore, the
results obtained using realistic forces are much closer to the data.
The agreement with the data might be improved if the self-energies
are determined including the full local momentum distributions (going beyond
a Fermi sphere) as it will be done in the near future.

\newpage
\large \bf Appendix
\normalsize \baselineskip 24pt
The usual density used in a non-relativistic framework is
defined as a sum over $N$ Gaussians describing the $N$ nucleons
\cite{Aichelin91},
i.e.
\begin{equation}
\rho_G (\vec r ,t)= \frac{1}{(2\pi L)^{3/2}}\,\sum_j^N\,\mbox{e}^{-(\vec r
- \vec r_j(t))^2/2L}.
\end{equation}
$L=1.08$ fm$^2$ fixes the width of a Gaussian.
Reasonable values for $\rho_G$ within the multi-time description of
RQMD are achieved by propagating all nucleons (forward or
backward) to the same
time coordinate in the frame in which $\rho_G$ has to be
determined, e.g. in the CMS.

In order to prove the reliability of the definitions for
the scalar density $\rho_S$ and the baryon density $\rho_B$ used
in eqs.(4) we determined the density profile of $\rho_G, \rho_S$
and $\rho_B$ of a Ca nucleus in its rest frame and for a boosted
situation (boosted to the CMS energy of a symmetric collision at
2 GeV/A) as well. The results of
these test calculations are shown in Fig. 5. In the rest frame of
the nucleus $\rho_G$ and $\rho_B$ coincide as it should be while
$\rho_S$ is smaller in the central region. In the boosted situation
$\rho_G$ and $\rho_B$ are enhanced by a $\gamma-$factor compared
to $\rho_S$. $\rho_G$ and $\rho_B$ show a quite similar shape in
the boosted situation which justifies that the definition used to determine
$\rho_B$ in RQMD simulations is reliable.
$\rho_G$ shows not the full Lorentz contraction because its
determination in RQMD simulations contains approximations as
mentioned above.
The reader should note that the density $\rho_G$
given at a special point $\vec r$ has half the width of the
so-called interaction density evaluated at the center of a Gaussian
$\vec r_i$ of a particle $i$. Hence, $\rho_B$ and $\rho_S$ as well
are also determined by using half the width of the corresponding
interaction density when evaluated at a special point in space time
as done in the calculations shown in Fig.5.
\\

\newpage
\large \bf Figure Captions
\normalsize \baselineskip 24pt\\
\vspace*{4mm}\\
{\bf Fig. 1:} Effective Couplings as functions of the scalar density.
Solid lines asign DBHF(Dirac Brueckner Hartree Fock)/BHF(Brueckner Hartree
Fock) calculations, dashed lines show the fit used in RQMD simulations.
QHDI denotes the values from the $\sigma-\omega$ model at the saturation
point.\\ \\
{\bf Fig. 2:} Root mean square radii of single Ca nuclei boosted to the CMS
energy of a symmetric collision at $1.8$ GeV/A as a function of time.
The different approaches DBHF, BHF and QHDI are explained in the caption of
figure 1. 'sat.' indicates for $\Gamma_{s}(\rho_{s})$ and
$\Gamma_{v}(\rho_{s})$ at the saturation density.\\ \\
{\bf Fig. 3:} Flow distribution versus rapidity at the final state (60fm/c)
of Ar+KCl collision at 1.8 GeV/A and $b\leq 2.3$ fm. The data are taken from
\cite{Danielewicz85}.\\ \\
{\bf Fig. 4:} Rapidity distribution at the final
state (60 fm/c) of Ar+KCl collisions at 1.8 GeV/A and $b\le 2.3$ fm.\\ \\
{\bf Fig. 5:} Usual density $\rho_G$, scalar density $\rho_S$ and baryon
density $\rho_B$
(definitions see text) for a single Ca nucleus
in the rest frame of the nucleus (left part) and in the CMS (right part).
\\ \\

\end{document}